\documentclass[a4paper,11pt]{article}
\pdfoutput=1 

\usepackage{jinstpub} 

\title{Wake field, impedance and collective instability}

\author
{Elias M\'etral
}

\affiliation
{CERN, Espl. des Particules 1,\\
1211 Meyrin, Switzerland}

\emailAdd{Elias.Metral@cern.ch}

\abstract{The first mention of the impedance concept appeared on November 1966 in the CERN internal report “Longitudinal instability of a coasting beam above transition, due to the action of lumped discontinuities” by V.G. Vaccaro. Then, a more general treatment of it appeared in February 1967 in the CERN yellow report “Longitudinal instabilities of azimuthally uniform beams in circular vacuum chambers of arbitrary electrical properties” by A.M. Sessler and V.G. Vaccaro. The concept of wake field came two years later, in 1969, in the paper “The wake field of an oscillating particle in the presence of conducting plates with resistive terminations at both ends” by A.G. Ruggiero and V.G. Vaccaro. This was the beginning of many studies, which took place over the last five decades, and today, impedances and wake fields continue to be an important field of activity, as concerns theory, simulation, bench and beam-based measurements. Building a reliable impedance or wake field model of a machine is the first necessary step to be able to evaluate the machine performance limitations, identify the main contributors in case an impedance reduction is required, and study the interaction with other mechanisms such as optics nonlinearities, transverse damper, noise, space charge, electron cloud, beam-beam (in a collider), etc. Beam collective instabilities, and their mitigation, cover a wide range of effects in particle accelerators and they have been the subjects of intense research. As the machines performance was pushed new mechanisms were revealed and nowadays the challenge consists in studying the interplays between all these intricate phenomena, as it is very often not possible to treat the different effects separately. With the increasing power of our computers this becomes easier but the need to continue and develop theories remains, to have a better understanding of the interplays between all these effects: the subject of impedance and beam instabilities in particle accelerators is far from being exhausted, as testified by the many new instability and stabilizing mechanisms which have been recently explained or discovered. Furthermore, in the context of the studies for possible future accelerators, some uncharted territories remain such as, for instance, the collective instabilities during the necessary ionization cooling for a muon collider.}


\proceeding{Special Issue: ICFA Beam Dynamics Newsletter No. 82}

\begin{document}
\maketitle
\flushbottom

\section{Introduction}
\label{intro}
As the beam intensity increases, the beam can no longer be considered as a collection of non-interacting single particles: in addition to the single-particle phenomena, collective effects become significant~\cite{RefChaoTextBook}~\cite{RefNgBook}~\cite{RefIEEE2016}~\cite{RefICFA-BDNL-2016}~\cite{EliasBenevento2017}~\cite{RefMCBI2019}. At low intensity a beam of charged particles moves around an accelerator under the Lorentz force produced by the external electromagnetic fields (from the guiding and focusing magnets, RF cavities, etc.). However, the charged particles also interact with themselves (leading to space charge effects) and with their environment, inducing charges and currents in the surrounding structures, which create electromagnetic fields called wake fields. In the ultra-relativistic limit, causality dictates that there can be no electromagnetic field in front of the beam, which explains the term wake. It is often useful to examine the frequency content of the wake field (a time domain quantity) by performing a Fourier transformation on it. This leads to the concept of impedance (a frequency domain quantity), which is a complex function of frequency.

If the wall of the beam pipe is perfectly conducting and smooth, a ring of negative charges (for positive charges travelling inside) is formed on the walls of the beam pipe where the electric field ends, and these induced charges travel at the same pace with the particles, creating the so- called image (or induced) current. But, if the wall of the beam pipe is not perfectly conducting or contains discontinuities, the movement of the induced charges will be slowed down, thus leaving electromagnetic fields (which are proportional to the beam intensity) mainly behind.

An ICFA mini-workshop on “Electromagnetic Wake Fields and Impedances in Particle Accelerators” was held in 2014 in Erice (Italy)~\cite{EriceWorkshop2014} to review the developments and main challenges in this field. They concerned the computation, simulation and measurement of the (resistive) wall effect for cylindrical and non cylindrical structures, any number of layers, any frequency, any beam velocity and any material property (conductivity, permittivity and permeability); the electromagnetic characterization of materials; the effect of the finite length of a structure; the computation and simulation of geometrical impedances for any frequency; the computation and simulation (in time and frequency domains) of all the transverse impedances needed to correctly describe the beam dynamics (i.e. the usual driving or dipolar wake, the detuning or quadrupolar wake, the angular wake, the constant and nonlinear terms, etc.); the issue of the wake function needed (inverse Fourier transform of the impedance, response to a delta-function) vs. the wake potential obtained from electromagnetic codes (i.e. response to a usually Gaussian pulse); the simulation of all the complexity of equipment like kickers, collimators and diagnostics structures; etc.

A second ICFA mini-workshop on "Impedances and Beam Instabilities in Particle  Accelerators" was held in 2017 in Benevento (Italy)~\cite{BeneventoWorkshop2017} to continue the work initiated with the 2014 ICFA mini-workshop. The aim of this event was first of all to provide an up-to-date review of the subject of beam coupling impedances (theory and modelling, simulation tools, bench measurements, beam based measurements). Besides, it was decided to widen its scope to include recent advances on theory and simulations of beam instabilities. It was also the occasion to celebrate V.G.~Vaccaro for the 50 years since the birth of the concepts of beam coupling impedance (and stability charts, better known nowadays as stability diagrams), to which he gave fundamental contributions unanimously recognised by the whole community. V.G. Vaccaro also received the Xie Jialin Prize for outstanding work in the accelerator field in 2019 at the IPAC2019 conference “For his pioneering studies on instabilities in particle beam physics, the introduction of the impedance concept in storage rings and, in the course of his academic career, for disseminating knowledge in accelerator physics throughout many generations of young scientists”.

Finally, to close the loop, a third workshop of the series, an ICFA mini-workshop on "Mitigation of Coherent Beam  Instabilities in Particle Accelerators", was held in 2019 in Zermatt (Switzerland)~\cite{RefMCBI2019}, which focused on all the mitigation methods for all the coherent beam instabilities, reviewing in detail the theories (and underlying assumptions), simulations and measurements on one hand, but on the other hand trying to compare the different mitigation methods (e.g. with respect to other effects such as beam lifetime) to provide the simplest and more robust solutions for the day-to-day operation of the machines.

This contribution will present an overview of the activities in the field of wake field, impedance and collective instabilities.

\section{Wakefield and impedance}
\label{WakefieldImpedance}
\subsection{Some historical considerations}
The first mention of the impedance concept appeared in 1966 in~\cite{VaccaroNov1966} and a more general treatment was published in 1967 in~\cite{VaccaroFeb1967}. The concept of wake field came two years later, in 1969, in~\cite{Vaccaro1969}. The impedance limits the performance of all the particle accelerators where the beam intensity (or beam brightness) is pushed, leading to beam instabilities and subsequent increased beam size and beam losses, and/or excessive (beam-induced RF) heating, which can deform or melt components or generate beam dumps. Each equipment of each accelerator has an impedance, which needs to be characterised and optimised. This impedance is usually estimated through theoretical analyses and/or numerical simulations before being measured through bench and/or beam-based measurements. Combining the impedances of all the equipment, a reliable impedance model of a machine can be built, which is a necessary step to be able to understand better the machine performance limitations, reduce the impedance of the main contributors and study the interplay with other mechanisms.

\subsection{Some theoretical aspects}
\begin{figure}[htbp]
\centering
\includegraphics[width=0.8\linewidth]{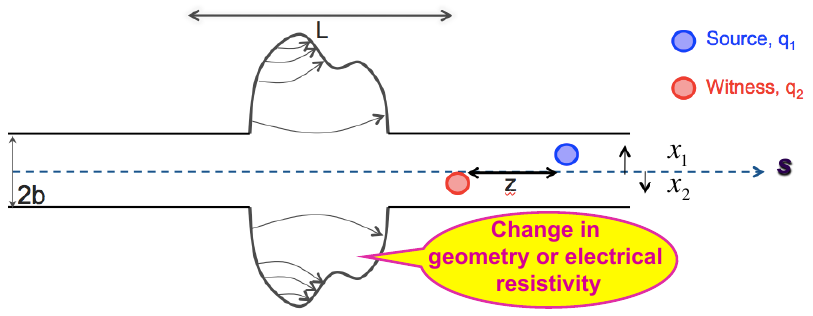}
\caption{Sketch of a vacuum chamber, which generates wake fields. Courtesy of G. Rumolo.}
\label{Picture1}
\end{figure}

What needs to be computed are the wake fields at a distance $z$ behind a source particle and their effects on the test or witness particles that compose the beam (see Fig.\ref{Picture1})~\cite{EliasBenevento2017}.

\begin{figure}[htbp]
\centering
\includegraphics[width=0.8\linewidth]{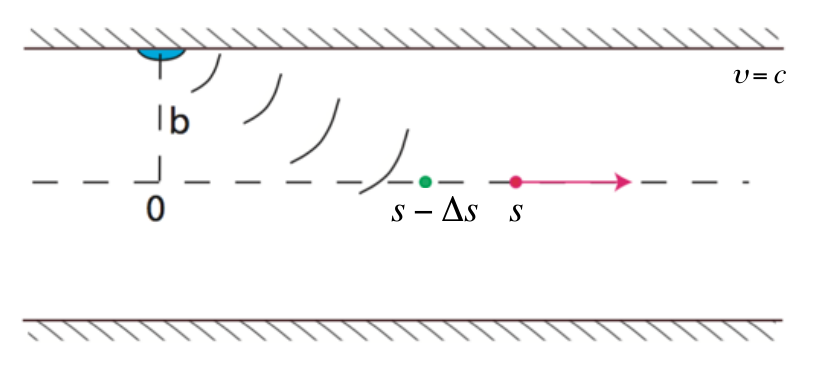}
\caption{A wall discontinuity located at $s = 0$ scatters the electromagnetic field of a relativistic particle. When the particle moves to location $s$, the scattered field arrives to point $s-\Delta s$. Courtesy of K. Bane and G. Stupakov~\cite{RefICFA-BDNL-2016}.}
\label{Picture2}
\end{figure}

For a particle moving along a straight line with the speed of light, due to causality, the electromagnetic field scattered by a discontinuity on the beam pipe does not affect the charges, which travel ahead of it. This field can only interact with the charges in the beam that are behind the particle, which generates the field. For short bunches, the time needed for the scattered fields to reach the beam on axis may not be negligible, and the interaction with this field may occur well downstream of the point where the field was generated. To find where the electromagnetic field produced by a leading charge reaches a trailing particle traveling at a distance $\Delta s$ behind the leading one, let’s assume that a discontinuity located on the surface of a pipe of radius $b$ at coordinate $s = 0$ is passed by the leading particle at time $t = 0$ with the speed of light $c$ (see Fig.\ref{Picture2}). It can be deduced from Fig.\ref{Picture2}, assuming $\Delta s << b$, that $s \approx b^{2}/(2 \Delta s)$. This distance $s$ is called the catch-up distance. Only after the leading charge has traveled that far away from the discontinuity, a particle at point $\Delta s$ behind it starts to feel the wake field generated by the discontinuity.

The computation of the wake fields is quite involved and two fundamental approximations are generally introduced: (i) the rigid-beam approximation (the beam traverses a piece of equipment rigidly, i.e. the wake field perturbation does not affect the motion of the beam during the traversal of the impedance. The distance of the test particle behind some source particle does not change) and (ii) the impulse approximation (as the test particle moves at a fixed velocity through a piece of equipment, the important quantity is the impulse, i.e. the integrated force, and not the force itself).

Starting from the four Maxwell equations for a particle in the beam and taking the rotational of the impulse, it can be shown that for a constant relativistic velocity factor $\beta$ (which does not need to be 1), a very general relation can be derived (without imposing any boundary conditions and considering only the two fundamental approximations), which is known as the Panofsky-Wenzel theorem~\cite{EliasBenevento2017}. Another important relation in the transverse plane can be obtained when $\beta=1$ (taking the divergence of the impulse).

Considering the case of a cylindrically symmetric chamber and as a source charge density (which can be decomposed in terms of multipole moments $m$) a macro-particle moving along the pipe with a transverse offset, the whole solution can be written for the azimuthal mode $m$, for $\beta = 1$, introducing the transverse wake function and its derivative which is called the longitudinal wake function. The latter describe the shock response (Green function) of the vacuum chamber environment to a $\delta$–function beam which carries an $m$th moment. The integrals of the Lorentz forces are called wake potentials (these are the convolutions of the wake functions with the beam distribution: here it is just a point charge). 

The Fourier transform of the wake function is called the impedance. As the conductivity, permittivity and permeability of a material depend in general on frequency, it is usually better (or easier) to treat the problem in the frequency domain, i.e. compute the impedance instead of the wake function. It is also easier to treat the case $\beta < 1$. Then, an inverse Fourier transform is applied to obtain the wake function in the time domain. Two important properties of impedances can be derived. The first is a consequence of the fact that the wake function is real, which leads to a relation giving the impedance for negative frequencies. The second is a consequence of Panofsky-Wenzel theorem, leading to a relation between the transverse and longitudinal impedances for the same azimuthal mode $m$.

Another interesting property of the impedances is the directional symmetry (Lorentz reciprocity theorem): the same impedance is obtained from both sides if the entrance and exit are the same. In the case of a cavity, an equivalent RLC circuit can be used (with three parameters which are the shunt impedance, the inductance and the capacity). In a real cavity, these three parameters cannot be separated easily and some other related parameters are used, which can be measured directly such as the resonance frequency, the quality factor (describing the width of the resonance) and the damping rate (of the wake). When the quality factor is low (and usually equal to 1), the resonator impedance is called broad-band, and this model was extensively used in the past in many analytical computations.

The situation is more involved in the case of non axi-symmetric structures (due in particular to the presence of the detuning wake field in addition to the usual driving wake field) and for $\beta < 1$, as in this case some electromagnetic fields also appear in front of the source particle. In the case of axi-symmetric structures, a current density with some azimuthal Fourier component creates electromagnetic fields with the same azimuthal Fourier component. In the case of non axi-symmetric structures, a current density with some azimuthal Fourier component may create an electromagnetic field with various different azimuthal Fourier components. Note that in the case $\beta < 1$, another detuning term is also found and a procedure to simulate or measure both the driving and detuning contributions can be deduced. In the time domain, using some time-domain electromagnetic codes like for instance CST Particle Studio, the driving and detuning contributions can be disentangled. The situation is more involved in the frequency domain, which is used for instance for impedance measurements on a bench. Two measurement techniques can be used to disentangle the transverse driving and detuning impedances, which are both important for the beam dynamics (this can also be simulated with codes like Ansoft-HFSS). The first uses two wires excited in opposite phase (to simulate a dipole), which yields the transverse driving impedance only. The second consists in measuring the longitudinal impedance, as a function of frequency, for different transverse offsets using a single displaced wire. The sum of the transverse driving and detuning impedances is then deduced applying the Panofsky-Wenzel theorem in the case of top/bottom and left/right symmetry. Subtracting finally the transverse driving impedance from the sum of the transverse driving and detuning impedances obtained from the one-wire measurement yields the detuning impedance only. If there is no top/bottom or left/right symmetry the situation is more involved and requires more measurements.
Finally, all the transverse impedances (driving and detuning) should be weighted by the betatron function at the location of the impedances, as this is what matters for the transverse beam dynamics.

\subsection{Numerical techniques}
Analytical computations are possible only if the structures are fairly simple. In practice this is often not the case and one has to rely on numerical techniques. First numerical wake field computations were performed in time domain by V. Balakin et al. in 1978~\cite{Balakin1978} and T. Weiland in 1980~\cite{Weiland1980}. As for highly relativistic bunches, due to causality, wake fields can catch up with trailing particles only after traveling the catch-up distance (see before), this motivated to compute wakes in linacs by using a mesh that moves together with the bunch: the moving mesh technique was introduced by K. Bane and T. Weiland in 1983~\cite{BaneWeiland1983}.

Nowadays many methods are available for beam coupling impedance simulations using either the Time Domain (TD) method or the Frequency Domain (FD) method (Eigenmode methods or methods based on beam excitation in FD) and the main electromagnetic codes currently used are ABCI, Ansys HFSS, CST Studio (MAFIA), GdfidL, ECHO2D, ACE3P, etc.

In TD, Finite Differences Time Domain (FDTD) and Finite Integration Technique (FIT) with leapfrog algorithm for the time stepping are used. More specialized techniques are the Boundary Element Method (TD-BEM), the Finite Volume method (FVTD), the Discontinuous Galerkin Finite Element Method (DG-FEM) or Implicit methods. The bunch length and the wake length are the two important parameters for TD impedance computations and the criterion for the time step is also referred to as the Courant-Friedrichs-Lewy (CFL) criterion. The TD simulations are suitable at medium and high frequencies, and particularly in perfectly conducting structures.

In FD, Eigenmode methods are used when high quality factor structures are under investigation and high accuracy is required.

The methods based on beam excitation in FD are well suited at low frequencies, where the CFL criterion poses a strong requirement on the time step. Due to the uncertainty principle, lower frequencies require computing longer wakes. As the time step is fixed by structure properties via the CFL criterion, this leads to the necessity to compute many time steps. The FD methods prevail for low frequency, low velocity of the beam and dispersively lossy materials.

\begin{figure}[htbp]
\centering
\includegraphics[width=0.6\linewidth]{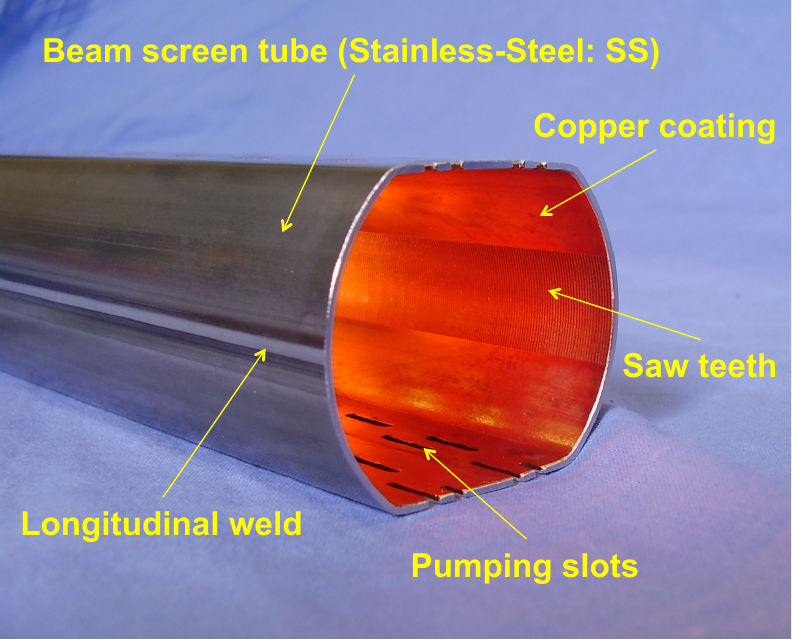}
\caption{Sketch of the LHC beam screen.}
\label{Picture3}
\end{figure}

The particularly difficult components to simulate are those, which combine elements of geometric wake fields and resistive elements (like tapered collimators or dielectric structures), the surface roughness and small random pumping slots (such as e.g. in Fig.\ref{Picture3}).

Finally, the electromagnetic properties of some materials (vs. frequency) are not well known and should be measured with precision before performing simulations to allow for reliable results.

\subsection{Analytical computations}
Analytical computations are usually used to compute the (resistive) wall impedance of multi-layered vacuum chambers, beam screens and collimators over a huge frequency range, as they are usually much faster and precise than simulation codes, which are facing several issues depending on the frequency range (number of mesh cells, etc.). Three formalisms are usually adopted: (1)~transmission-line, (2) field matching and (3) mode matching.

For a cylindrical beam pipe or two parallel plates, with any number of layers, any beam velocity and any electric conductivity, permeability and permittivity, the IW2D code was derived using field matching~\cite{IW2D}, which is valid when the length of the structure is (much) larger than the beam pipe radius (or half gap in the case of two parallel plates)~\cite{RefIEEE2016}.

\begin{figure}[htbp]
\centering
\includegraphics[width=0.8\linewidth]{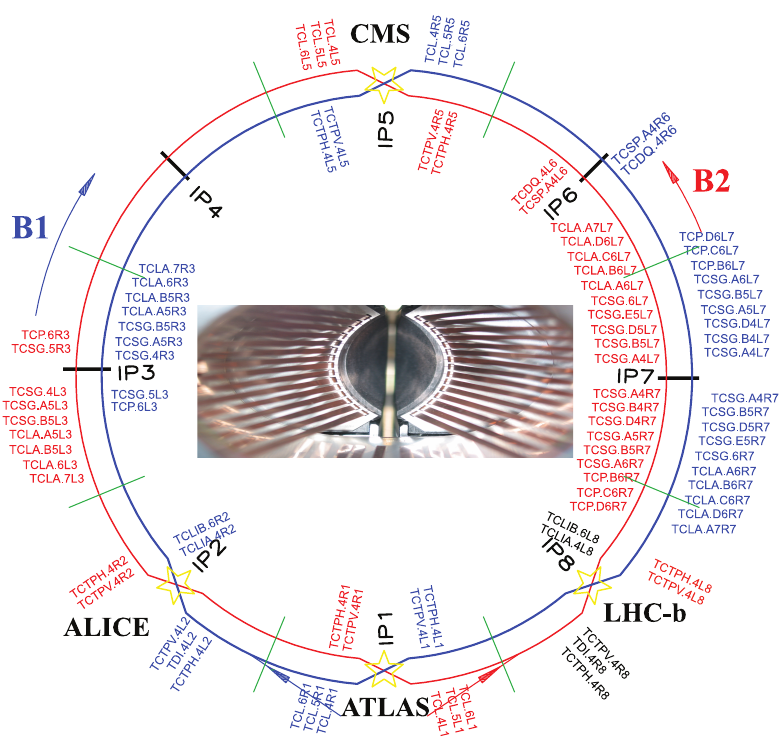}
\caption{The numerous collimators of the LHC, whose distance to the beam is of few mm. Courtesy of S.~Redaelli.}
\label{Picture4}
\end{figure}

\begin{figure}[htbp]
\centering
\includegraphics[width=0.8\linewidth]{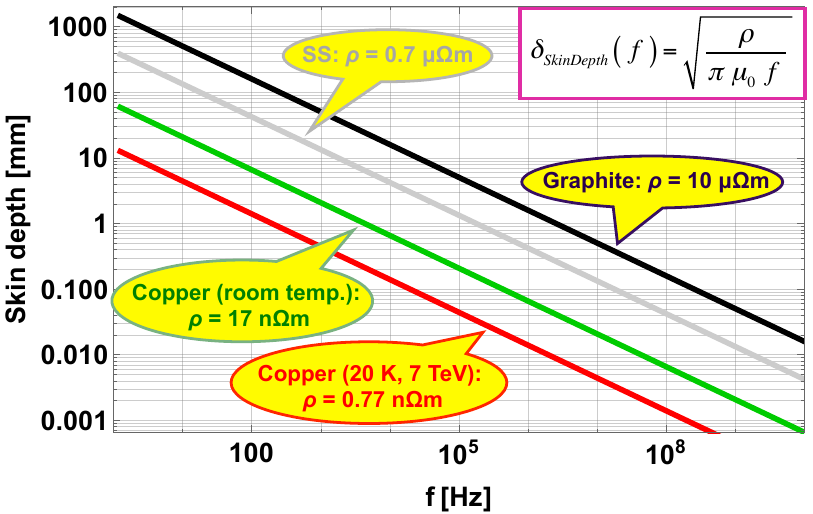}
\caption{Skin depth vs. frequency for different resistivities: stainless steel, graphite and copper (at room and cryogenic temperatures). Here, $\rho$ is the electrical resistivity and $\mu_{0}$ is the vacuum permeability.}
\label{Picture5}
\end{figure}

\begin{figure}[htbp]
\centering
\includegraphics[width=0.8\linewidth]{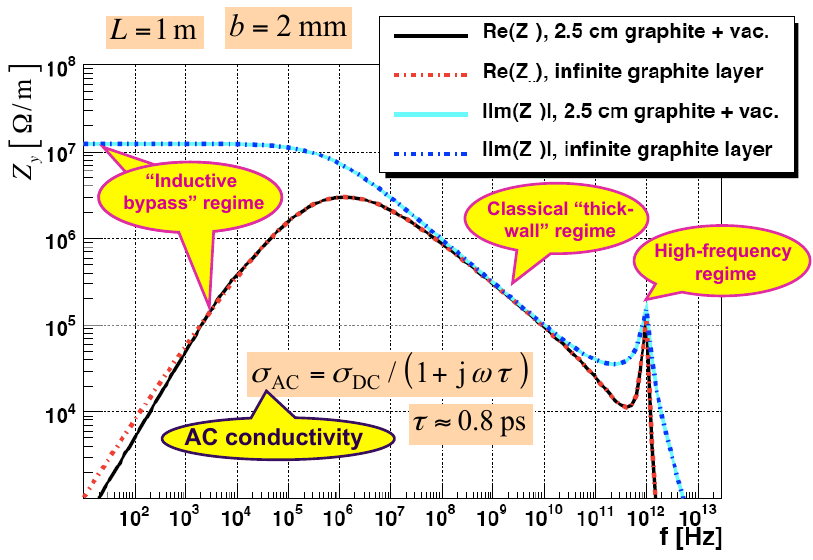}
\caption{Transverse wall impedance for the case of a (round) LHC collimator. Here, $\sigma$ is the electrical conductivity and $\tau$ is the relaxation time.}
\label{Picture6}
\end{figure}

\begin{figure}[htbp]
\centering
\includegraphics[width=0.8\linewidth]{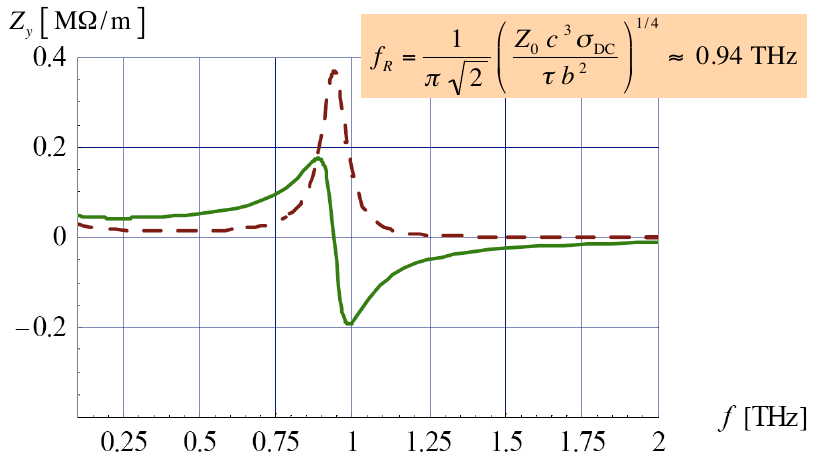}
\caption{Zoom of the third (high-frequency) regime of Fig. 6: real (in brown) and imaginary (in green) parts of the transverse impedance. Here, $Z_0$ is the vacuum impedance.}
\label{Picture7}
\end{figure}

In the CERN LHC, the wall impedance of the numerous collimators (see Fig.\ref{Picture4}) represents a significant fraction of the total machine impedance. The important parameters are the beam pipe radius (or half gap in the case of two parallel plates), the thickness of the different layers and the skin depth, which is plotted vs. frequency in Fig.\ref{Picture5} for three different materials. Assuming for simplicity first the case of a (round) LHC collimator, the transverse wall impedance is represented in Fig.\ref{Picture6}, exhibiting three regimes of frequencies: (1)~low-frequency or inductive bypass regime; (2)~intermediate-frequency or classical thick-
wall regime and (3) high-frequency regime.
Before discussing in detail the first two regimes, which are of interest for the LHC, let’s have a closer look at the third (high-frequency) regime, which is zoomed in Fig.\ref{Picture7}. A resonance is clearly revealed, whose physical interpretation was provided by K. Bane~\cite{Bane1991}. The beam/wall interaction can be thought of occurring in two parts: first the beam loses energy to the high frequency resonator and then, on a longer scale, this energy is absorbed by the walls.

\begin{figure}[htbp]
\centering
\includegraphics[width=0.8\linewidth]{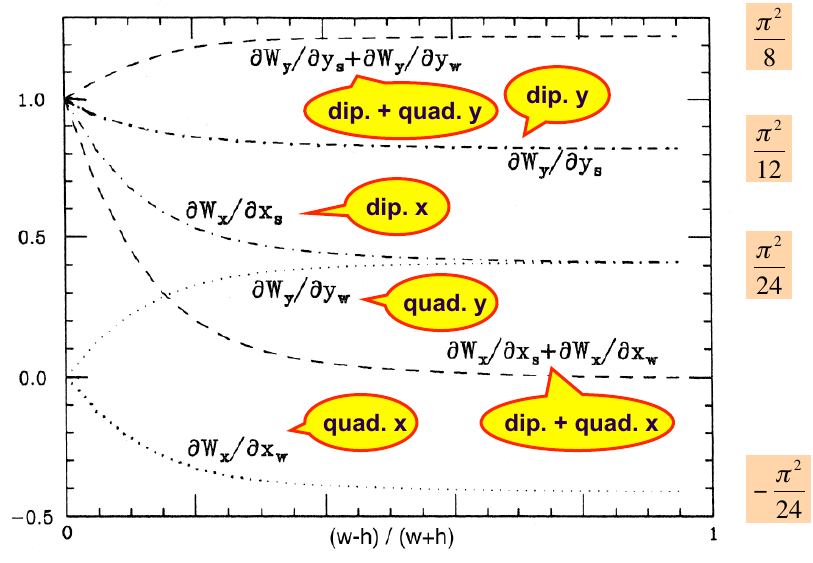}
\caption{Yokoya form factors for dipolar and quadrupolar impedances in resistive elliptical pipes (compared to the circular case)~\cite{Yokoya1993}. The height of the elliptical beam pipe is $h$ and the width is $w$ (the case where $w = h$ corresponds to the circular case).}
\label{Picture8}
\end{figure}

\begin{figure}[htbp]
\centering
\includegraphics[width=0.8\linewidth]{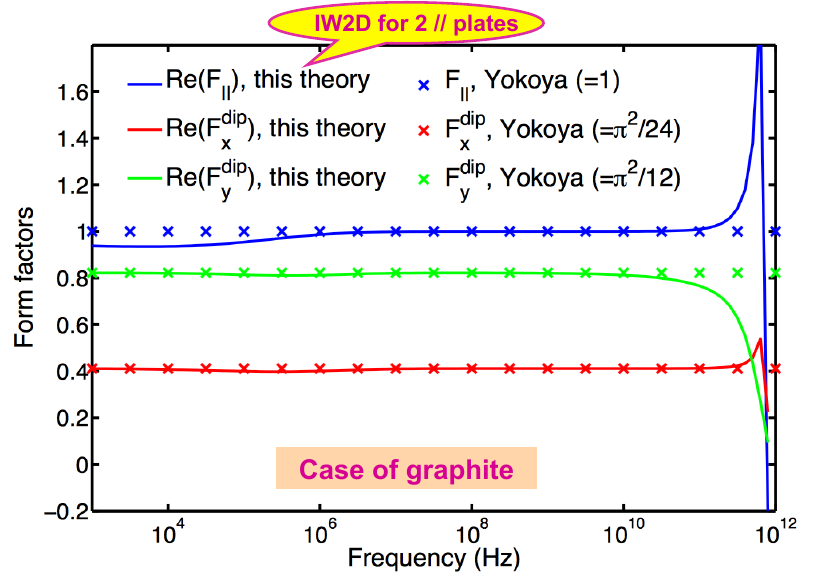}
\caption{Generalised form factors (compared to the circular case) from the IW2D code (mentioned as “this theory”) for the case of graphite~\cite{GeneralizedFormFactors2010}.}
\label{Picture9}
\end{figure}

\begin{figure}[htbp]
\centering
\includegraphics[width=0.8\linewidth]{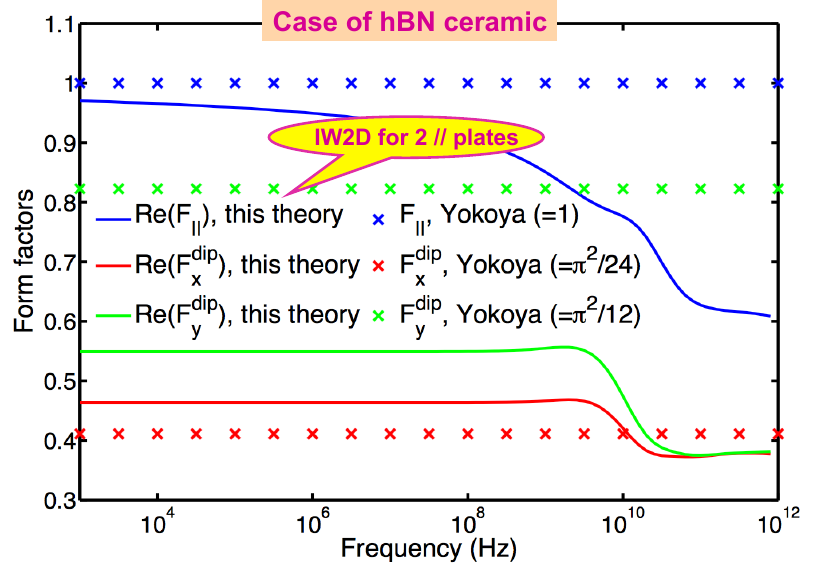}
\caption{Generalised form factors (compared to the circular case) from the IW2D code (mentioned as “this theory”) for the case of hBN ceramic~\cite{GeneralizedFormFactors2010}.}
\label{Picture10}
\end{figure}

In the case of resistive elliptical beam pipes, Yokoya form factors~\cite{Yokoya1993} for the dipolar and quadrupolar impedances are usually used (see Fig.\ref{Picture8}): these form factors are numbers to be multiplied to the results obtained with the circular geometry (with the height of the elliptical beam pipe $h$ equal to the radius $b$ of the circular beam pipe). However, it should be reminded that several assumptions were made as concerns both the frequency and the material. Generalised form factors have been deduced from IW2D for two parallel plates compared to the circular case and two examples are shown in Figs.\ref{Picture9} and \ref{Picture10}.

\subsection{Transverse wall impedance and coatings}
\begin{figure}[htbp]
\centering
\includegraphics[width=0.8\linewidth]{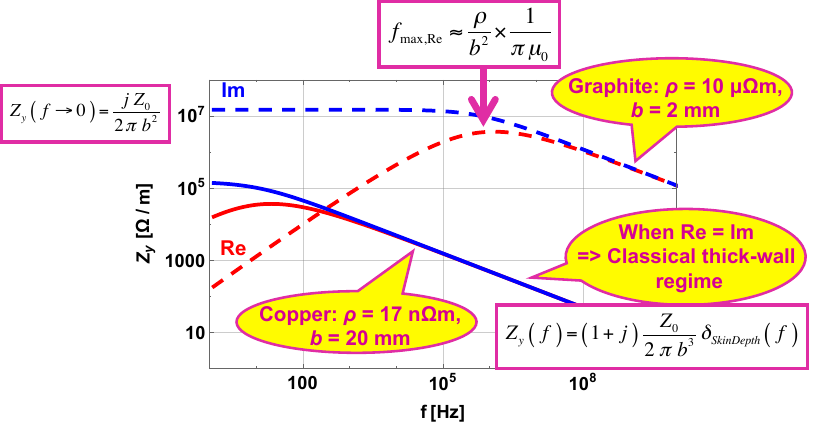}
\caption{Transverse wall impedance assuming a round beam pipe with a length of 1 m and only one layer going to infinity.}
\label{Picture11}
\end{figure}

Assuming a round beam pipe with a length of 1 m with only one layer going to infinity, the first two frequency regimes are depicted in Fig.\ref{Picture11} for two different beam pipe radii and conductivities.

\begin{figure}[htbp]
\centering
\includegraphics[width=0.6\linewidth]{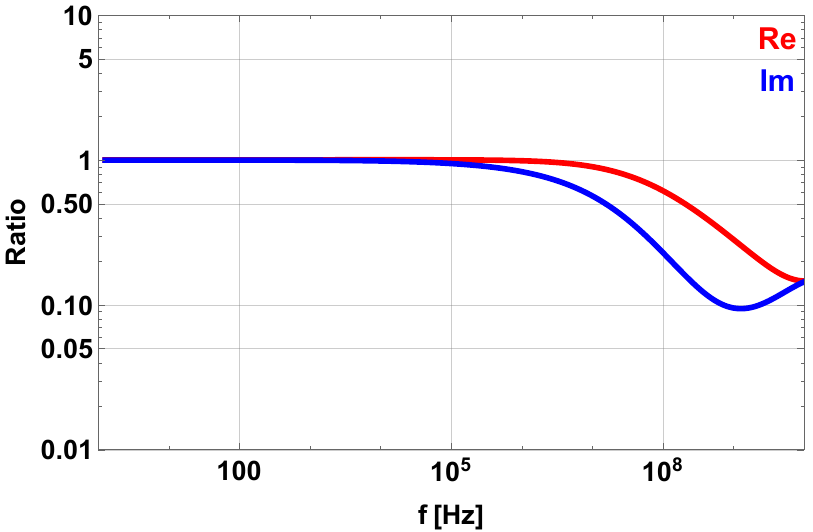}
\caption{Effect of a copper coating (at room temperature) inside a round beam pipe with a length of 1~m, one layer of stainless steel going to infinity and a radius of 20 mm. Ratio of the impedance (real and imaginary parts) with the coating of thickness 1 $\mu$m to the impedance without the coating.}
\label{Picture12}
\end{figure}

\begin{figure}[htbp]
\centering
\includegraphics[width=0.6\linewidth]{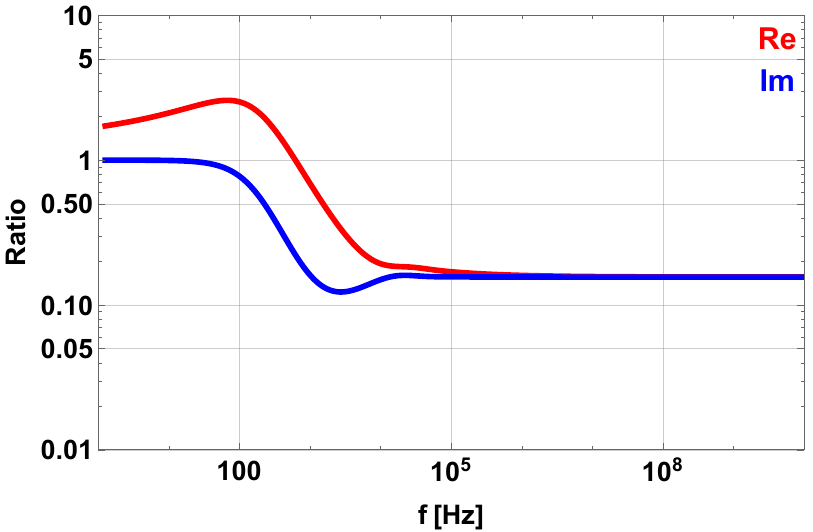}
\caption{Effect of a copper coating (at room temperature) inside a round beam pipe with a length of 1~m, one layer of stainless steel going to infinity and a radius of 20 mm. Ratio of the impedance (real and imaginary parts) with the coating of thickness 1000 $\mu$m to the impedance without the coating.}
\label{Picture13}
\end{figure}

The effect of a copper coating inside a round beam pipe with a length of 1 m, one layer of stainless steel going to infinity and a radius of 20 mm, is represented in Figs.\ref{Picture12} and \ref{Picture13}. It is shown that the imaginary part of the impedance is always reduced while a too high thickness of the coating can considerably increase the real part of the impedance at low frequencies (as the better conductor keeps the induced current closer to the beam). 

\begin{figure}[htbp]
\centering
\includegraphics[width=0.6\linewidth]{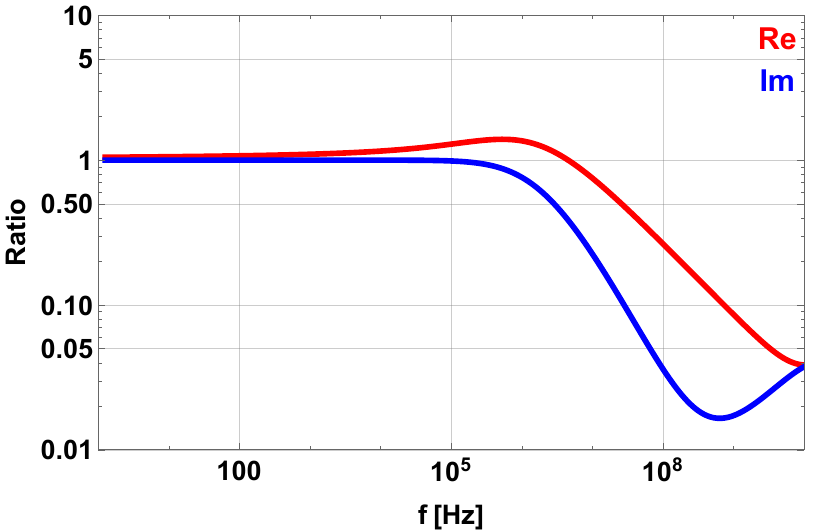}
\caption{Effect of a copper coating (at room temperature) inside a round beam pipe with a length of 1 m, one layer of graphite going to infinity and a radius of 2 mm. Ratio of the impedance (real and imaginary parts) with the coating of thickness 1 $\mu$m to the impedance without the coating.}
\label{Picture14}
\end{figure}

\begin{figure}[htbp]
\centering
\includegraphics[width=0.6\linewidth]{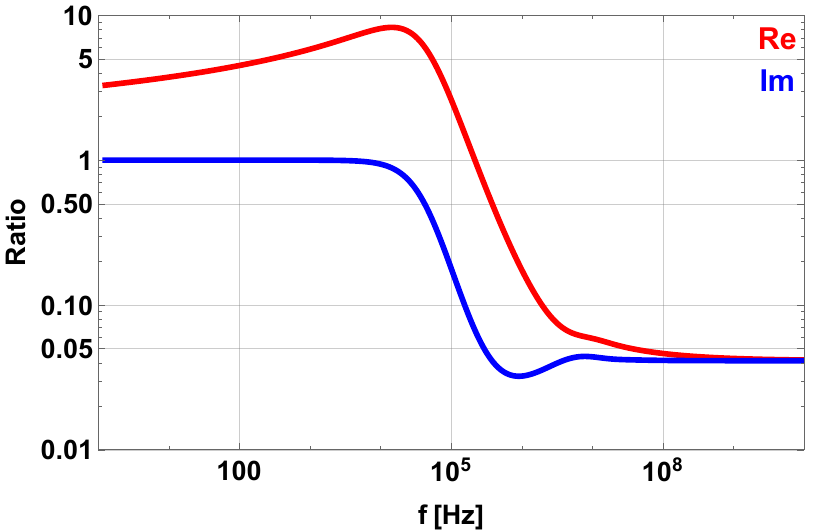}
\caption{Effect of a copper coating (at room temperature) inside a round beam pipe with a length of 1 m, one layer of graphite going to infinity and a radius of 2 mm. Ratio of the impedance (real and imaginary parts) with the coating of thickness 50 $\mu$m to the impedance without the coating.}
\label{Picture15}
\end{figure}

The effect of a copper coating inside a round beam pipe with a length of 1 m, one layer of graphite going to infinity and a radius of 2 mm, is represented in Figs.\ref{Picture14} and \ref{Picture15}. As before (but with an effect, which is amplified), it is shown that the imaginary part of the impedance is always reduced while a too high thickness of the coating can considerably increase the real part of the impedance at low frequencies (as the better conductor keeps the induced current closer to the beam). 

\begin{figure}[htbp]
\centering
\includegraphics[width=0.6\linewidth]{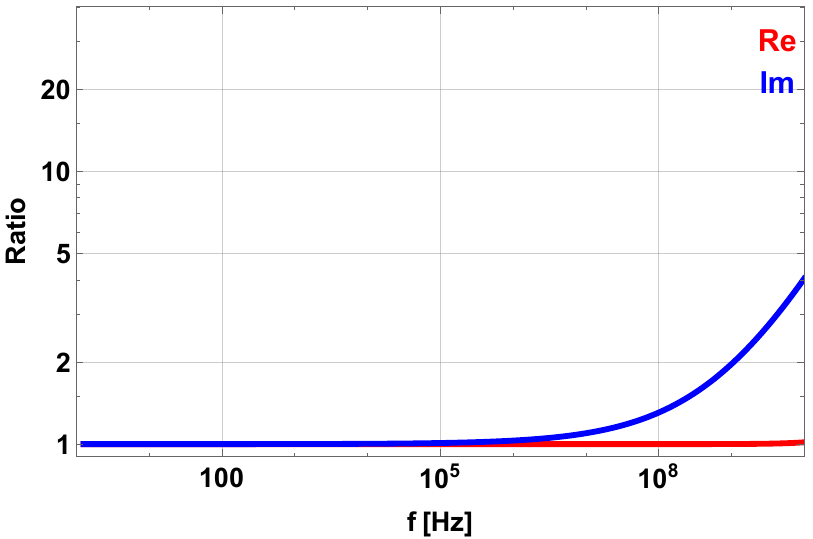}
\caption{Effect of a graphite coating inside a round beam pipe with a length of 1 m, one layer of copper (at room temperature) going to infinity and a radius of 20 mm. Ratio of the impedance (real and imaginary parts) with the coating of thickness 1 $\mu$m to the impedance without the coating.}
\label{Picture16}
\end{figure}

\begin{figure}[htbp]
\centering
\includegraphics[width=0.6\linewidth]{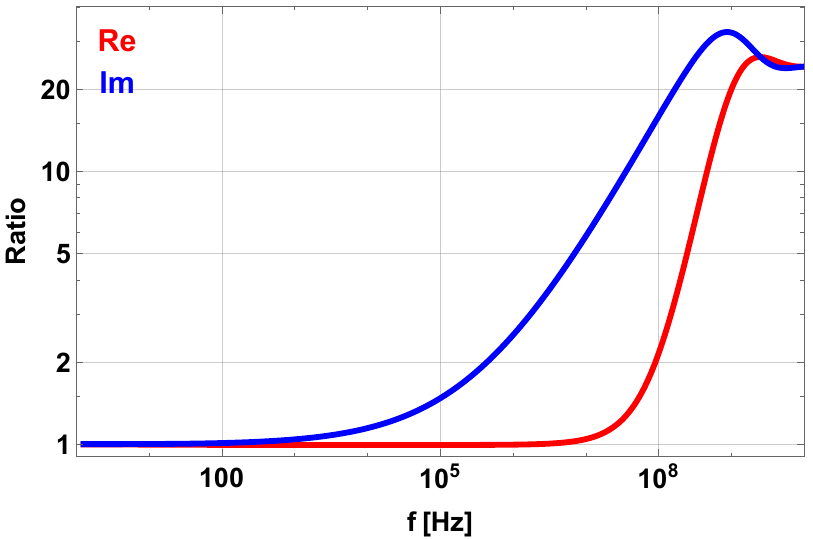}
\caption{Effect of a graphite coating inside a round beam pipe with a length of 1 m, one layer of copper (at room temperature) going to infinity and a radius of 20 mm. Ratio of the impedance (real and imaginary parts) with the coating of thickness 50 $\mu$m to the impedance without the coating.}
\label{Picture17}
\end{figure}

The effect of a graphite coating (e.g. as could be used to reduce the secondary emission yield and relevant electron cloud effects) inside a round beam pipe with a length of 1 m, one layer of copper (at room temperature) going to infinity and a radius of 20 mm, is represented in Figs.\ref{Picture16} and \ref{Picture17}. In this case, it is shown that if the coating thickness is sufficiently small, the real part of the impedance does not change but only the imaginary part increases at high frequency. For larger coating thicknesses, even the real part of the impedance will be significantly higher for high frequencies, which could have detrimental effects for beam stability.

In summary, even if the impedance is more than fifty years old, most of the particle accelerators do not have a sufficiently precise impedance model and there are still challenges for the future, such as e.g. with the new surface treatments, which need to be implemented to fight against electron cloud.

\subsection{Extension of the impedance concept and uncharted new territories}
Based on the success of the concept of wake field or impedance to describe the interaction between a beam and its surrounding, several extensions of the impedance concept appeared over the years for space charge, electron cloud and CSR (Coherent Synchrotron Radiation). These simplified models can bring a lot of very interesting information but it is always very important to analyse in detail all the approximations made to be sure that the model remains sufficiently precise and useful, depending on the subject under study.

In the context of the studies for possible future accelerators, some uncharted territories remain such as the collective instabilities during the necessary ionization cooling for a muon collider. This subject is of paramount importance because such mechanism could jeopardize the generation of high brightness muon beams through ionization cooling. The knowledge of collective instabilities that could arise from the interaction of the beam with electromagnetic wake fields propagating in matter (absorbers, gas-filled RF cavities, etc.) as well as with the pair of charges generated by ionization is practically non-existing. These effects will have to be studied in detail in the coming years to check that a muon collider is a viable option for the future, and the first step will be to elaborate models which describe the electromagnetic wake fields and impedances generated by a beam passing through matter. 

\section{Collective instability}
\label{Collective instability}
\subsection{Some historical considerations}
\begin{figure}[htbp]
\centering
\includegraphics[width=0.6\linewidth]{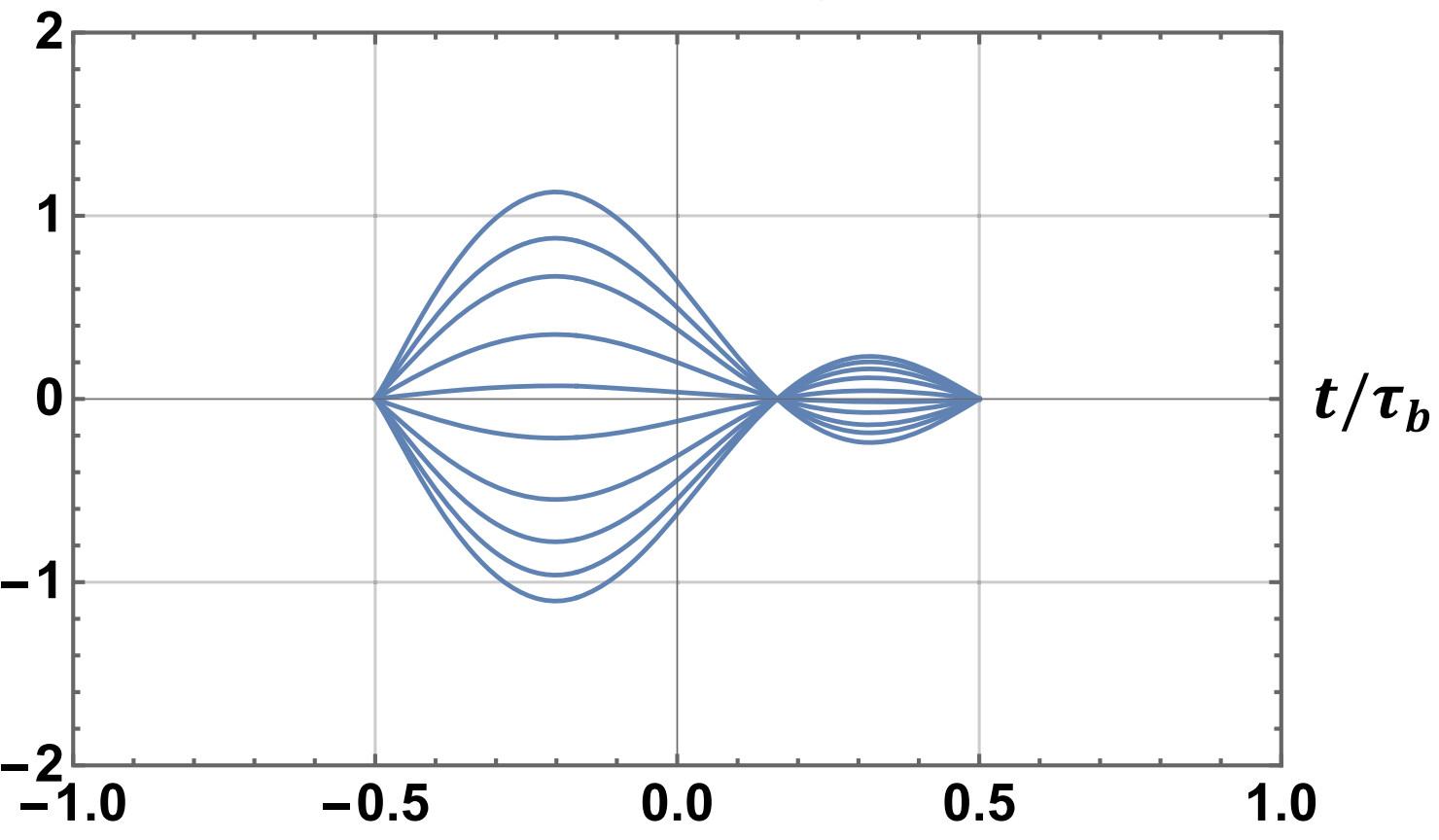}
\caption{Example of head-dominated intrabunch motion (with the traces of several turns superimposed) in the case of transverse mode coupling between modes 0 and -1 at the start of the instability~\cite{RefIntraBunch}). Here, $\tau_b$ is the full bunch length and $t$ is the time. }
\label{Picture18}
\end{figure}

Impedance-driven (but not only) coherent beam instabilities are usually studied analytically with the linearized Vlasov equation, ending up with an eigenvalue system to solve. The motion of the beam is then described by a superposition of modes rather than a collection of individual particles. This mode analysis was initially developed by F. Sacherer in 1972~\cite{Sacherer1972} and extended by others~\cite{RefChaoTextBook}. The eigenvalues lead to the beam oscillation mode-frequency shifts, leading in particular to the longitudinal mode coupling instability (LMCI) and transverse mode coupling instability (TMCI)~\cite{MetralMigliorati2020}. These beam oscillation mode-frequency shifts can be directly measured or simulated for the lowest modes and in the absence of nonlinearities but this becomes difficult for high-order modes and/or in the presence of important nonlinearities. The eigenvectors lead to the intrabunch motion, which had been explained in the past in particular by Laclare~\cite{LaclareCAS1987} for independent modes, i.e. at sufficiently low intensity (leading to symmetric intrabunch signals between the head and the tail, called standing-wave patterns), while the intrabunch motion in the presence of several modes was only recently discussed, opening the possibility to explain analytically asymmetric intrabunch signals as already simulated with macroparticle tracking codes and observed in our particle accelerators~\cite{RefIntraBunch}. The coupling between two standing waves is a traveling wave, which is another way to see that the bunch is in the TMCI regime. The TMCI is head dominated close to the intensity threshold (see Fig.\ref{Picture18}~\cite{RefIntraBunch}) and the signal moves from the head to the tail when the bunch intensity is increased. Better characterizing an instability is the first step before trying to find appropriate mitigation measures and push the performance of a particle accelerator. The evolution of the intrabunch motion with intensity is a fundamental observable with high-intensity high-brightness beams. A lot of work remains to be done in this direction comparing the predictions and beam-based measurements vs. intensity to better understand all the observed instabilities. 

\subsection{Impedance-induced instabilities}
\begin{figure}[htbp]
\centering
\includegraphics[width=0.8\linewidth]{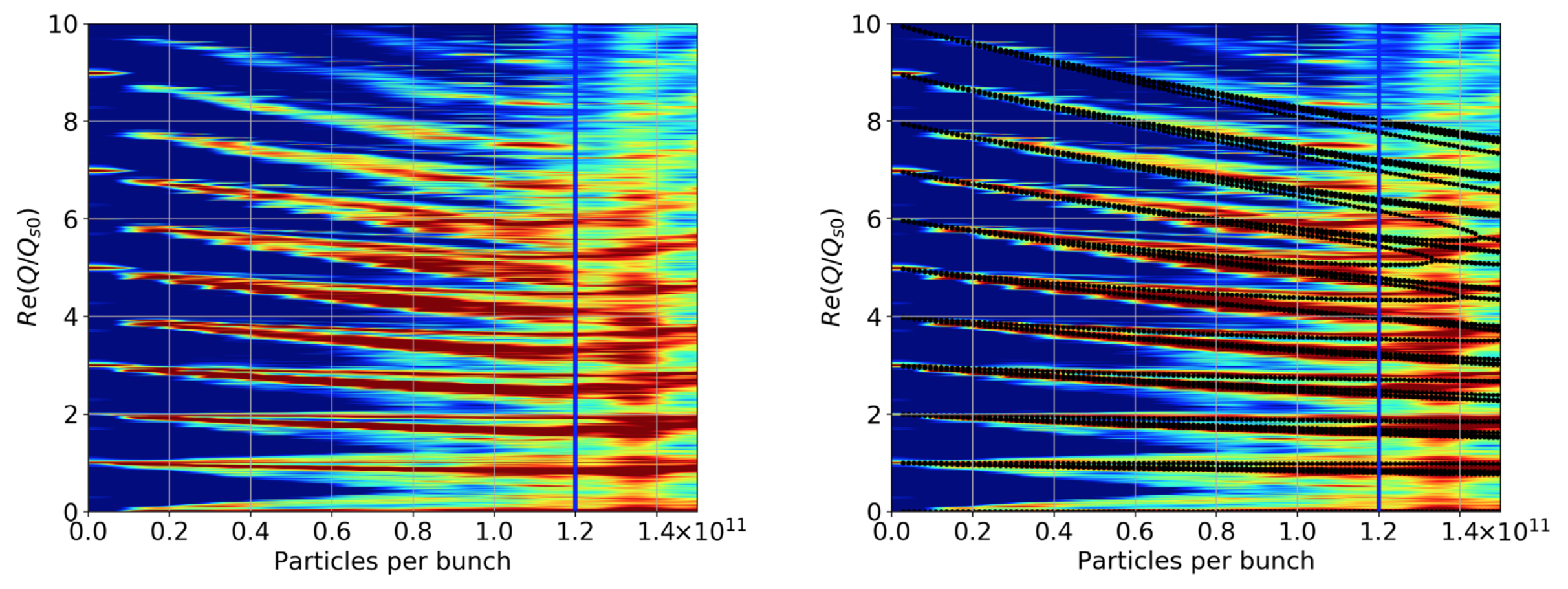}
\caption{Comparison of the real part of the normalized mode-frequency shift between the SBSC tracking code and the GALACLIC Vlasov solver (black dots in the right-hand side) for the case of a broad-band impedance~\cite{MetralMigliorati2020}. Here, $Q_{s0}$ is the low-intensity small-amplitude synchrotron tune.}
\label{Picture19}
\end{figure}

\begin{figure}[htbp]
\centering
\includegraphics[width=0.8\linewidth]{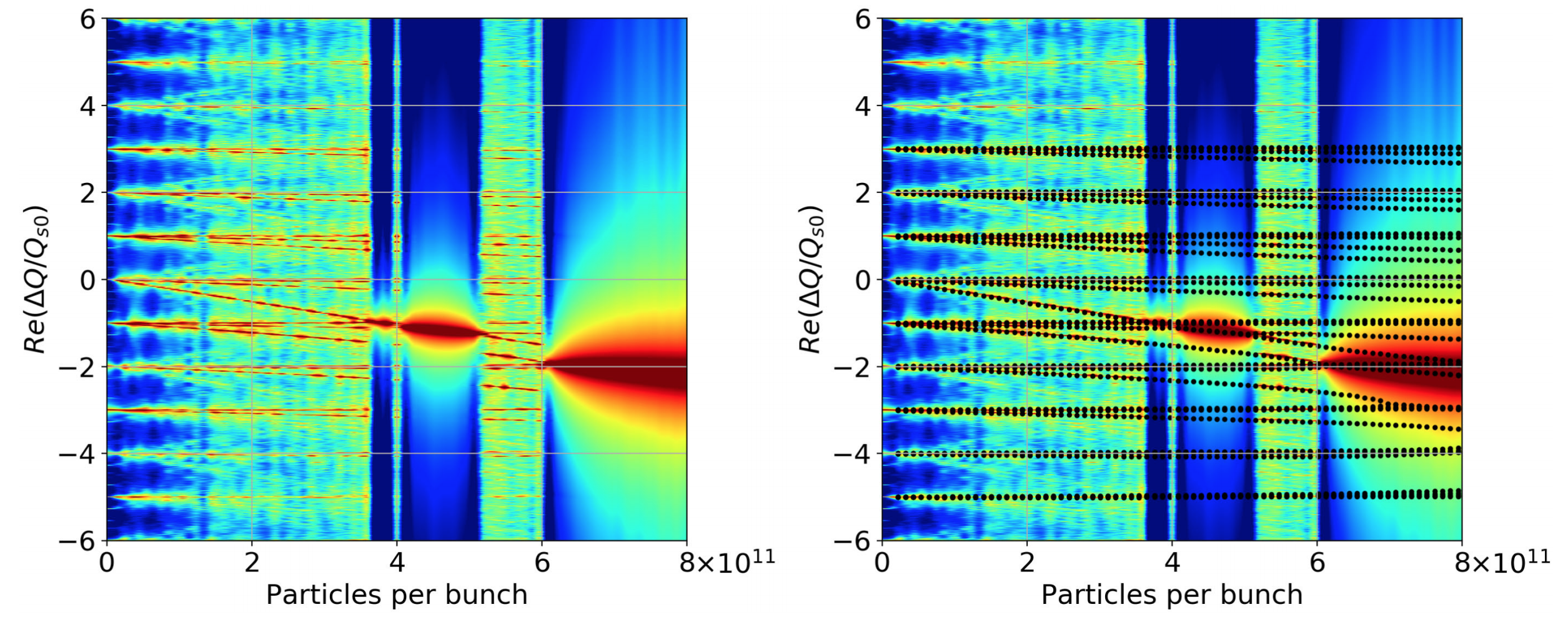}
\caption{Comparison of the real part of the normalized mode-frequency shift between the PyHEADTAIL tracking code and the GALACTIC Vlasov solver (black dots in the right-hand side) for the case of a broad-band impedance~\cite{MetralMigliorati2020}. Here, $Q_{s0}$ is the low-intensity small-amplitude synchrotron tune.}
\label{Picture20}
\end{figure}

\begin{figure}[htbp]
\centering
\includegraphics[width=0.6\linewidth]{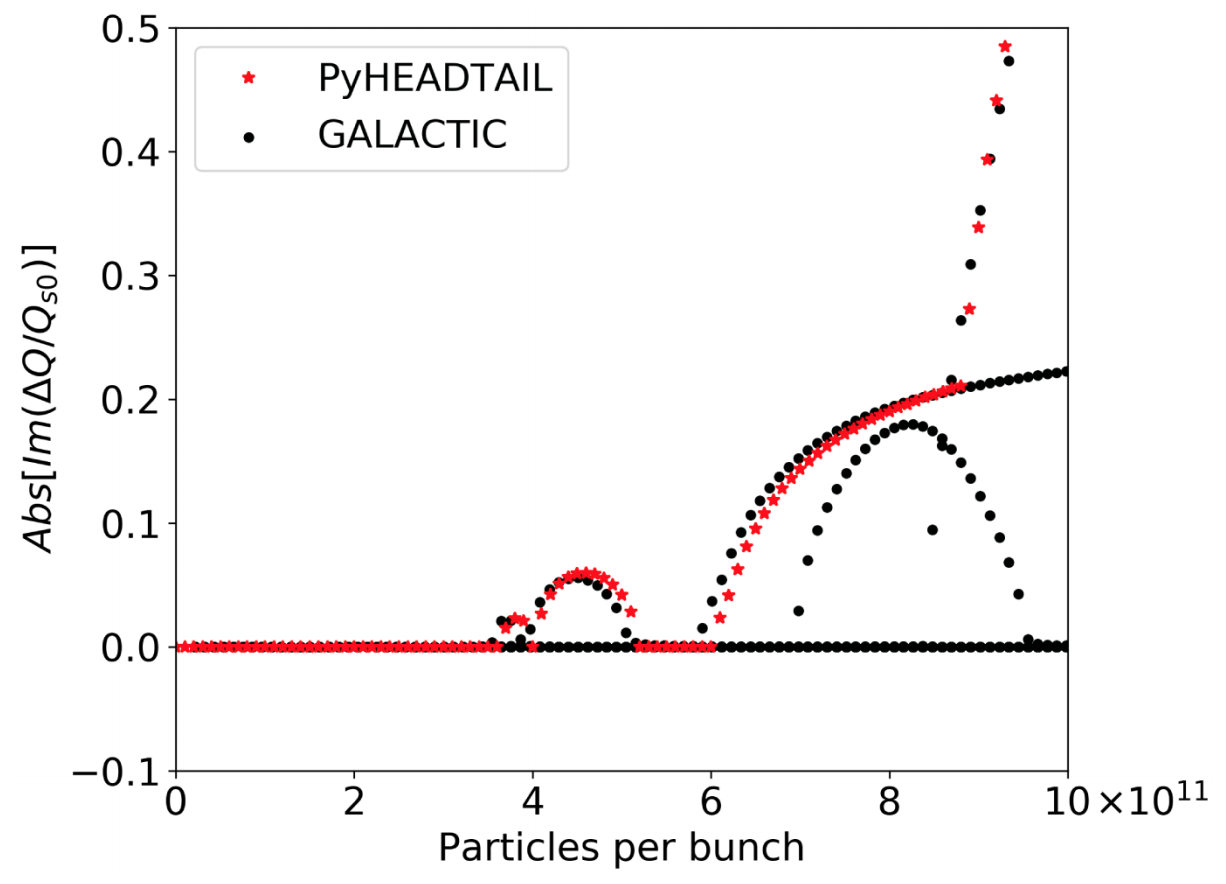}
\caption{Growth rates of the transverse mode coupling instability as a function of the bunch intensity given by the PyHEADTAIL tracking code and the GALACTIC Vlasov solver (black dots)~\cite{MetralMigliorati2020}. Here, $Q_{s0}$ is the low-intensity small-amplitude synchrotron tune. Observe that GALACTIC gives the
growth rates of several unstable modes.}
\label{Picture21}
\end{figure}

Several Vlasov solvers have been developed over the years in both transverse and longitudinal planes such as GALACTIC and GALACLIC, which have been benchmarked in some regimes as discussed in~\cite{MetralMigliorati2020} (see Figs.\ref{Picture19}, \ref{Picture20} and \ref{Picture21}). It is worth mentioning that in GALACTIC the effect of a transverse damper was also added, as it was already the case with two other transverse Vlasov solvers widely used by the community, NHTVS~\cite{NHTS} and DELPHI~\cite{DELPHI}. These three Vlasov solvers are similar (with some pros and cons) as concerns the subject of this paper as they are solving the same equation but using different formalisms, and it was checked in the past that on few benchmark cases the same results were obtained. It is also worth mentioning that
DELPHI uses the same approach as the famous MOSES code from Y.H. Chin, which has been used for several decades (since 1988),
where the Sacherer integral equation is solved using a
decomposition over Laguerre polynomials of the radial
functions~\cite{MOSES}. Furthermore, the MOSES code had been already successfully benchmarked against the HEADTAIL macroparticle tracking code in~\cite{PHDthesisOfBenoitSalvant}. Other approaches to solve the Vlasov equation in the longitudinal plane, valid in particular for electron machines, can be found in~\cite{Cai2011}~\cite{Schonfeldt2017}~\cite{Warnock2006}.

\begin{figure}[htbp]
\centering
\includegraphics[width=0.6\linewidth]{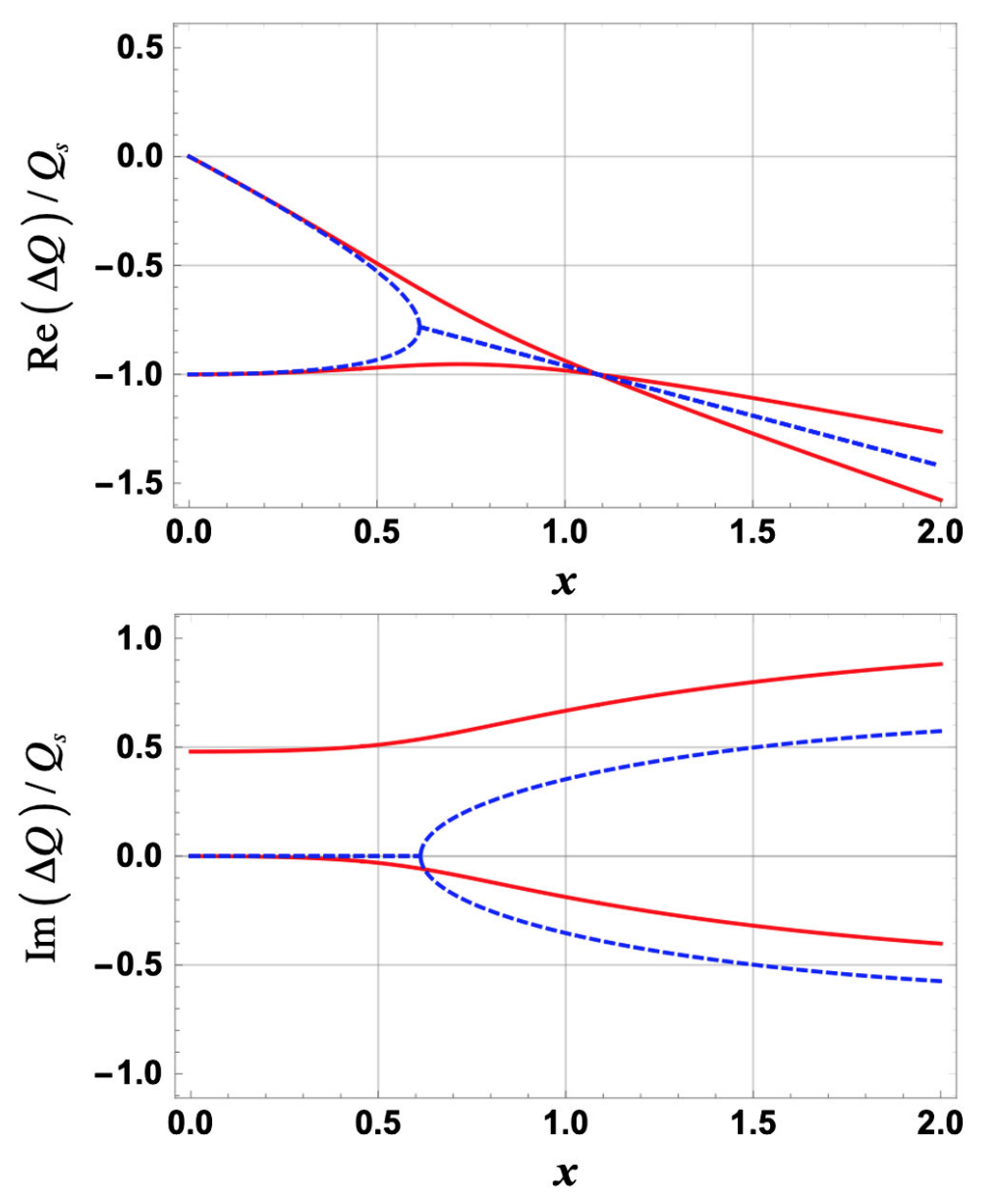}
\caption{Comparison between the TMCI instability without a resistive transverse damper (dashed blue line) and the ITSR instability in the presence of a resistive transverse damper (full red line)~\cite{RefITSR}. The parameter $x$ is a normalized parameter proportional to the bunch intensity and $Q_{s}$ is the low-intensity small-amplitude synchrotron tune.}
\label{Picture22}
\end{figure}

A lot of progress has been made over the years with the tracking codes where many effects can be added together but the current challenge lies in the analytical development of the corresponding Vlasov solvers to have a better understanding of the different physical mechanisms involved (in addition to the continuation of the further development of the numerical tools). Over the last few years, the main focus was to add the effect of a transverse damper. A resistive transverse damper is often needed in particle accelerators operating with many bunches and it
is usually very efficient as it can considerably reduce the necessary amount of nonlinearities needed to
reach beam stability through Landau damping (which provides a natural stabilizing mechanism against collective instabilities if the particles of the beam have a small spread in their incoherent frequencies). However, a resistive transverse damper also destabilizes the single-bunch motion below the TMCI intensity threshold (for zero chromaticity), introducing a new kind of
instability, which has been called ITSR instability (for imaginary tune split and repulsion)~\cite{RefITSR} (see Fig.\ref{Picture22}).

\begin{figure}[htbp]
\centering
\includegraphics[width=0.6\linewidth]{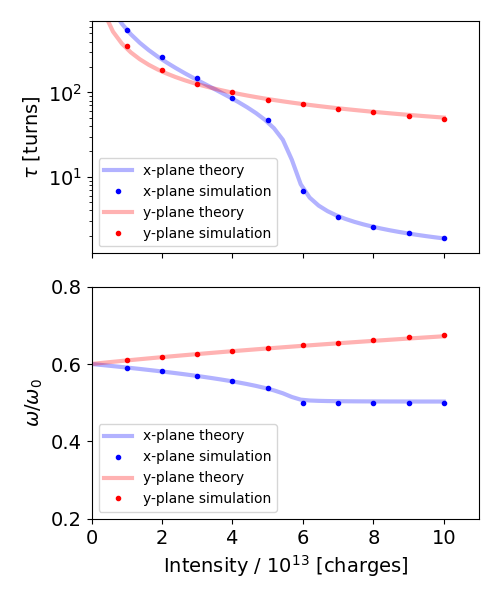}
\caption{PyHEADTAIL simulations (with dots) compared to the new theory
accounting for coupling between fast and slow waves (with full
lines) for the horizontal (blue) and the vertical (red) planes. The
rise time, at the top, and the normalized (by the revolution frequency) frequency shift, at the
bottom, are shown~\cite{RefNB-MM-EM-2}. }
\label{Picture23}
\end{figure}

Another important effect which needs to be carefully taken into account to correctly describe the beam dynamics with asymmetric impedances, is the detuning impedance. Introducing it for coasting beams, a new instability was revealed, the mode coupling instability for coasting beams, which was never discussed or described in the past~\cite{RefNB-MM-EM-1}~\cite{RefNB-MM-EM-2}~\cite{PRABcommentBurovLebedev}~\cite{ReplyToPRABcommentBurovLebedev} (see Fig.\ref{Picture23}), and which revealed a plane exchange of the most critical instability vs. intensity (which was one of the important findings of the study). The detuning impedance has been also recently added in the Vlasov analysis with bunched beams (using also some simulation results), leading to an excellent agreement with the PyHEADTAIL macroparticle tracking code~\cite{RefIadarolaPRABecloud} (see also next section). The current challenge is to develop a fully self-consistent Vlasov solver for generalized impedances (without and with Landau damping).

Many other instability mechanisms have been identified over the past years by studying the interplays with other mechanisms, such as the coupled head-tail instability due to the destabilising effect of linear coupling~\cite{RefLeeCarver} (and references therein), the destabilising effect of mode-coupling of colliding beams (beam-beam and impedance)~\cite{RefSimonWModeCouplingCollidingBeams}, the destabilising effect of the interplay between Landau octupoles and beam-beam~\cite{RefBuffatStabilityDiagrams}, the destabilising effect of noise~\cite{RefSondreFuruseth}, the destabilising effect of space charge~\cite{RefBuffatPRABtmciAndSC} (and references therein), etc.

\subsection{Electron cloud instabilities}
The electron cloud build-up and related instabilities have been generally studied in the past with tracking codes due to the complexity of the mechanisms involved. Some useful simplified analytical models were however developed, extending for instance the concept of impedance for electron cloud. Recently, a linearized method to study the transverse instabilities due to electron clouds was presented, solving the linearized Vlasov equation, taking into account both the cloud dipolar and quadrupolar forces (which was an important missing ingredient in some past analyses), that can be obtained from quick single-pass numerical simulations. The method was successfully benchmarked against macroparticle tracking simulations based on the same characterization of the electron cloud and the results were also compared against the conventional simulations based on the Particle-In-Cell method. The effect of transverse nonlinearities due to the electron cloud,
which are neglected by the linearized method, were also analyzed~\cite{RefIadarolaPRABecloud}.

\subsection{Multi-species instabilities}
The presence of ions and electrons from gas ionization, photoemission or secondary emission is unavoidable in the vacuum chambers of high intensity accelerators and storage rings. Under certain conditions, these ions and electrons can accumulate and drive the beams unstable. A recent status report was provided in~\cite{RefLottaMatMCBI2019} (and references therein) where the mechanisms behind and the main conditions for ion and electron accumulation in bunched beams were summarized. The characteristics of the induced instabilities, as well as common modelling techniques and mitigation strategies were also reviewed. Electron and ion instabilities have been observed in several machines. Electron cloud is present in several operating machines. Ion instabilities are currently observed mainly under vacuum degradation, but may become more prevalent in future machines with higher beam brightness. 

Furthermore, it is worth mentioning that an instability mechanism relying on the interplay between electrons and ions may also occur under exceptional conditions. If large amounts of electrons and ions are generated at each bunch passage, e.g. if very high gas densities occur in the beam pipe, one can expect the two species to have a significant impact on each other’s dynamics. This could result in a collective behaviour involving the two species that is qualitatively different from the behaviour of either species on its own. Events falling under this category occurred in the LHC during operation in 2017 and 2018. The recurring events, which were characterized by very fast beam instabilities accompanied by unusual beam losses in a certain machine location, would inevitably lead to beam dumps. The instabilities are thought to have been caused by a very high local gas density, predicted by beam loss rates to 10$^{18}$–10$^{22}$ m$^{-3}$, depending on the longitudinal extent of the gas. These transient pressure bumps were generated by the beam-induced phase transition of macroparticles of frozen air, present due to an accidental inlet of air during the preceding machine cool-down. Observations of large positive tune shifts and fast intrabunch motion suggest that large electron densities were present during the events. However, electron cloud simulations with high gas densities could not reproduce the observations. To model the instability, electron cloud simulation tools were extended with multi-cloud capabilities to model both the build-up and beam stability in the presence of electrons and ions simultaneously. For high gas densities, the simulations show a significant impact of the field from the ion population on the electron dynamics, confirming that, at high concentrations, the two species must be modelled together, which is the current challenging task.

\subsection{Mitigation of coherent beam instabilities}
Beam instabilities, and their mitigation, cover a  wide range of effects in particle accelerators and they have been the subjects of intense research for several decades~\cite{RefMCBI2019}. The beam coupling impedance is the first cause of coherent beam instabilities but many other mechanisms are important  to  take  into  account  to  quantitatively  describe the observed instability mechanisms and thresholds of our particle  accelerators. And not  only  all  these mechanisms have  to  be  understood  separately,  but all the  possible interplays  between  the  different  phenomena  need  to  be analyzed  in  detail,  including  the  beam coupling  impedance  (both driving  and  detuning ones  in  the  transverse planes), the  linear  and  nonlinear  chromaticity, the  transverse  damper (including  a detailed  description vs. frequency of  the  gain, the bandwidth and the noise), the Landau octupoles (and other intrinsic nonlinearities), space charge, beam-beam (with both long-range and head-on effects for colliders), electron cloud or/and ions, linear coupling strength, tune separation between the transverse  planes (bunch by bunch), tune split between the two beams (bunch by bunch, for colliders), transverse beam  separation  between  the  two  beams (for colliders), noise, etc.

For existing machines, trying to push the machine performance usually requires a detailed impedance model and a systematic analysis to identify the main contributors and reduce their impedance, which requires a lot of time and resources. In case of instabilities due to additional electrons or ions, all the methods to try and reduce/suppress the latter should be put in place (it is worth noting for instance that nanostructuring of material surfaces by laser ablation is a well-established science and manufacturing with more than 25 years of experience). In many machines, longitudinal and transverse  feedbacks exist and realistic modellings of them need to be included in the beam stability analyses, as they considerably modify  both  coupled-bunch  and  single-bunch  motions (it  is worth noting that in some machines, many  feedbacks are used: 35 feedback loops are used for instance in the CERN PS machine for the LHC-type beams!). Feedbacks are  working  very  well but  they  cannot, at the moment (this might change in the future if we succeed to reach the necessary bandwidth), damp all types of instabilities. And in some cases, as already discussed before, some destabilising  effects  from  feedbacks  can also  be  observed. Therefore, the  next  step  is  to optimise the machine linear and nonlinear optics (tunes, linear and nonlinear  chromaticity, amplitude detuning, linear coupling, transition energy, etc.) and the RF knobs (such as RF voltage or controlled longitudinal emittance) and rely on Landau damping, which requires the introduction of a spread in the incoherent frequencies to absorb the energy of the coherent motion. However, a trade-off between coherent beam stability and single-particle stability (i.e. dynamic aperture) needs to be found and all the sources of Landau  damping  should  be  considered  and  studied carefully together: some  interplays  can  be  beneficial and  some others  can be detrimental. Furthermore,  it  was  recently found that the transverse stability diagram (deduced from the analysis  of  Landau  damping  for independent  coherent beam modes) can significantly evolve with time in the presence of noise, which could then be detrimental to the long-term beam stability. Landau octupoles are usually used to provide Landau damping in the transverse plane but some other proposals have been made, such as using beam-beam (long-range and/or head-on) in colliders, or an RF quadrupole (which has a similar effect as the second-order chromaticity, providing longitudinal-to-transverse Landau damping)~\cite{RFQgrudiev}~\cite{RFQSchenk1}~\cite{RFQSchenk2}~\cite{RFQSchenk3}, or an electron lens~\cite{ElectronLensesPRL} (which is doing something similar as the beam-beam head-on in non-collider rings). All these different methods should be analysed in detail and compared to find the best mitigation measure depending on the relevant conditions of operation of a particular particle accelerator. During the 2019 MCBI workshop in Zermatt~\cite{RefMCBI2019}, I.~Hofmann, the chair of the ICFA Beam Dynamics panel~\cite{ICFA-BD-Panel}, motivated the whole community to work more on the effect of Landau damping, as there is still a lot to be done on Landau damping and its possible loss, looking in more detail to theories, simulations and measurements (e.g. with Beam Transfer Function, or, as recently proposed, using an anti-damper as a controlled source of impedance~\cite{AntipovPRL}). 

For future machines, it is recommended to try and integrate all the above aspects already in the design stage, i.e. reduce as much as possible the impedance and electron cloud (surface) effects (in close collaboration with all the equipment  groups) and optimise the optics design by including already from the beginning all the collective effects.

Finally,  the knowledge  transfer  and  collaboration  between experts and operations are key to ensure that all the teams are moving in the same direction to produce stable beams. Indeed, if one takes as an example the CERN Proton Synchrotron: on November 24th, 1959, this first strong-focusing proton synchrotron ever built accelerated a proton beam to a kinetic energy of 24 GeV, becoming for a brief period the world's highest energy particle accelerator. Since then, its intensity per pulse has been increased by more than three orders of magnitude and it is currently part of the LHC accelerator complex. This would not have been possible without a sound initial design and without all the efforts and the ingenuity of generations of accelerator physicists, engineers, operators, and technicians, allowing for maintainability, flexibility, and versatility of this unique accelerator.

\section{Conclusions and outlook}
\label{outlook}
In his famous textbook of Physics of Collective Beam Instabilities in High Energy Accelerators~\cite{RefChaoTextBook}, A. Chao said: ``Over the years, I have learned and been fascinated by this subject, and it is this fascination that I would like to share with the reader''. I would like to thank him very much for his ``bible'' which came perfectly on time for my PHD thesis and all my future activities~\cite{EliasForChaoSymposium}. I have also learned and been fascinated by this subject, and it is this fascination that I would like to share with the next generation. 

I hope that this contribution convinced you that despite all the excellent work performed over the last decades, the subject of wake field, impedance and beam instabilities in particle accelerators is far from being exhausted and it remains a cutting-edge field in beam physics. Many new instability and stabilizing mechanisms have been recently explained or discovered and many challenges remain to be tackled with some uncharted territories such as, for instance, the collective instabilities during the necessary ionization cooling for a muon collider. This subject is of paramount importance because such mechanism could jeopardize the generation of high brightness muon beams. Based on the recent 2019 MCBI workshop~\cite{RefMCBI2019}, the community seems motivated to exchange experience and join efforts to advance further. The amount of open questions, the continuing progress recorded on different fronts and the promising outlook of many studies in terms of development and search for solutions fully legitimate the quest to pursue this series of workshops, which proved to be extremely useful for the whole community.


\end{document}